\documentclass[reprint,amsmath,amssymb,aps]{revtex4-2}
\usepackage{natbib} 
\usepackage{graphicx}
\usepackage{caption,subcaption}
\begin{document} 
	\title{Distinct viscoelastic scaling for isostatic spring networks of the same fractal dimension}

	\author{Aikaterini Karakoulaki} 
	\affiliation{School of Physics and Astronomy, University of Leeds, LS2 9JT, United Kingdom. }
	
	\author{David Head}
	\affiliation{School of Computing, University of Leeds, LS2 9JT, United Kingdom. }
		
	\begin{abstract}
		Fractal structure emerges spontaneously from the chemical cross\-linking of monomers into hydrogels, and has been directly linked to power law visco\-elasticity at the gel transition, as recently demonstrated for isostatic (marginally--rigid) spring networks based on the Sierpinski triangle. Here we generalize the Sierpinski triangle generation rules to produce 4 fractals, all with the same dimension $d_{\rm f}=\log 3/\log 2$, with the Sierpinski triangle being one case. We show that spring networks derived from these fractals are all isostatic, but exhibit one of two distinct exponents for their power--law viscoelasticity. We conclude that, even for networks with fixed connectivity, power--law viscoelasticity cannot generally be a function of the fractal dimension alone.
	\end{abstract} 
	
	%
	%

	\maketitle

	%
	%
	\section*{Introduction} 

	A broad range of soft materials including proteins~\cite{AufderhorstRoberts2020,Hughes2022,Ikeda1999}, polymers~\cite{Chambon1987,Adolf1990}, and colloids~\cite{Zaccone2014,Aime2018}, self--assemble into load bearing, disordered configurations when immersed within a suitable reaction environment. The microscopic structure of such gels are largely frozen in at the point of kinetic arrest, when the first spanning cluster emerges and diffusion becomes suppressed, so that the properties of the ultimate gel can be related to properties at the gel transition~\cite{Martin1991}. Early polymer gel experiments revealed that the viscoelastic spectrum $G^{*}(\omega)=G^{\prime}(\omega)+iG^{\prime\prime}(\omega)$, with $G^{\prime}(\omega)$ the storage (in--phase) and $G^{\prime\prime}(\omega)$ the loss (out-of-phase) linear response to a sinusoidal shear of frequency $\omega$~\cite{BarnesBook}, demonstrated power law scaling $G^{\prime}(\omega)\propto G^{\prime\prime}(\omega)\propto \omega^{\Delta}$, extending down to the lowest frequencies attainable for systems close to the gel point~\cite{Chambon1987,Ding1988}.
	
	The range of possible mechanisms for power--law viscoelasticity is broad; however, most involve some form of micro\-scale structural relaxation~\cite{Hwang2016,Song2022}, which does not occur in chemically--cross\-linked protein hydrogels that nonetheless exhibit $G^{*}(\omega)\propto \omega^{\Delta}$~\cite{AufderhorstRoberts2020,Hughes2021}. When the network connectivity is thus fixed, any power--law response must derive from a broad range of network structural features. Indeed, early theoretical approaches attempted to relate $\Delta$ to the fractal dimension that emerges as the monomers aggregate into the network~\cite{Hughes2021,Jungblut2019}. These theories tended to fall into one of two camps based on the assumptions made~\cite{Martin1988,Martin1989,Muthukumar1985}, but always produced expressions for $\Delta$ that depended only on the fractal dimension of the gel $d_{\rm f}$, the spatial embedding dimension~$d$, and the hydrodynamic conditions (i.e. Rouse versus Zimm)~\cite{DoiEdwards}.
	
	Recently, a class of $d=2$ dimensional spring networks was introduced that exhibited a geometry that was fractal up to a controllable length scale~\cite{SpringNetworks}. The fractal was the Sierpinski triangle, previously used to investigate the rigidity properties of systems with scale--invariant correlations~\cite{Machlus2021}, but with modified generation rules so that the corresponding spring networks were marginally rigid. This means their connectivity was precisely the minimum required for the system to respond as a solid, and were therefore analogous to molecular systems at the gel transition. Moreover, the connectivity was fixed, as for chemically--crosslinked hydrogels, so structural rearrangements were impossible. Numerical calculation of the linear viscoelastic response, following the scheme of Yucht et al.~\cite{Yucht2013}, demonstrated a power--law regime $G^{\prime}\propto G^{\prime\prime}\propto\omega^{\Delta}$ extending down to frequencies that depending on the maximum extent of fractal microstructure, but the exponent remained constant, $\Delta\approx0.22$. A simple scaling argument was presented that predicted a value for $\Delta$ based on the fractal dimension of the network, which was in good agreement with the measured value. However, as these spring networks were based on the same fractal --- the Sierpinksi triangle (or gasket) with $d_{\rm f}=\log 3/\log 2$~\cite{FractalGeometry} --- it was not possible to test this argument for different fractals.	
	
	Here we devise a class of fractals that generalizes the Sierpinski triangle, following a similar generalization of square (or `box') fractals~\cite{BoxFractal}. Four fractals are produced, including the standard Sierpinski triangle as one case, and all have the same fractal dimension $d_{\rm f}=\log 3/\log 2$. They are also all marginally rigid, in that the connectivity of their respective spring networks, as quantified by the mean number of springs connected to a node $\langle z\rangle$, approaches the marginal rigidity value $2d=4$ for large systems. However, when the viscoelastic spectrum was calculated, two values of $\Delta$ were measured amongst the four fractals, $\Delta\approx0.29$ in addition to the previously measured $\Delta\approx0.22$, despite the fractal dimension $d_{\rm f}$, the embedding dimension $d$, and the hydrodynamic conditions (Rouse), being the same for all systems considered. We conclude that a full theory for the gel--point viscoelastic exponent $\Delta$ for systems with fixed connectivity, must require more information as input than just the fractal dimension $d_{\rm f}$ of the gelled network, in contrast to the assumptions of previous approaches.

	%
	%
	\section*{Methods}
	
	\subsection*{Generalization of the Sierpinski triangle}
\begin{figure}[htbp]
		\begin{center}
			\includegraphics[width=8.6cm]{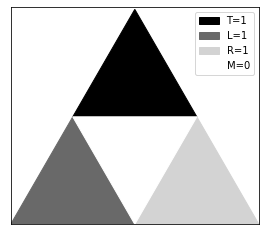}
		\end{center} 
		\caption{The Sierpisnki triangle for 
			one level of recursion with sub-triangles having different shade depending on their type (T,L,R,M).}
		\label{fig:1}
	\end{figure}

	The standard Sierpinski triangle (or gasket) is generated by repeatedly replacing an equilateral triangle by three triangles of half the size at the vertices, leaving an empty inverted triangle of the same size in the middle~\cite{FractalGeometry}. This is one possible generation rule can be described by the four variables $T$ (top), $L$ (left), $R$ (right) and $M$ (middle), that each take the value 0 or 1 depending on if the corresponding triangle should exist at the next level of recursion or not; see Fig.~\ref{fig:1}. Thus, the rules just described are given by the set
	\begin{equation}
		T=1, L=1, R=1, M=0 \:.
	\end{equation}
	Each smaller triangle is then subdivided according to the same rules, recursively generating the Sierpinski triangle.

	In order to generalize the Sierpinski triangle, the rules of the next level should depend on the previous level triangle's state, where the state can be either 0 (empty triangle) or 1 (filled triangle). This process is based on the 'box fractal' iteration rules~\cite{BoxFractal}. In total, eight variables need to be defined. The first set of four variables represent the rules for the next recursive level for triangles with state 1 in the previous level,
	\begin{equation}
		T1, L1, R1, M1 \:.
	\end{equation}
	\newline 
	The second four rules represent give the same information for when the triangle at the previous level of recursion did not exist, that is, had a state of 0,
	\begin{equation}
		T0, L0, R0, M0\:.
	\end{equation} 
	\newline
	 Again these variables can take values 0 or 1 depending on the state of the respective triangle in the next level of recursion. In the case of the Sierpinski triangle the rules are
	\begin{eqnarray}
		T1=L1=R1=1,M1&=&0,\\
		T0=L0=R0=M0&=&0
	\end{eqnarray}
	since the middle triangle must stay empty (state 0) for all recursive levels. 

	\begin{figure}[htbp]
		\begin{center}
			\includegraphics[width=8.6cm]{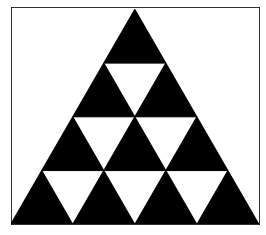}
		\end{center} 
		\caption{ If the rules of the Sierpisnki triangle  were to stay the same except of M0 (changed from 0 to 1), this object would be produced for 2 levels of recursion. }
		\label{fig:2}
	\end{figure}

	An example is given in Fig.~\ref{fig:2}, which shows that for every triangle with state 0, in this case the middle triangle, the second set of rules (T0,L0,R0,M0) are followed in the next level, and for triangles with state 1 the first set of rules are used, (T1,L1,R1,M1). Since the eight variables defined above can take values 0 or 1, there are $2^8=256$ combinations, or in other words generation rules. In order to analyse the generated objects based on their geometry, we developed code that returns all the triangles' coordinates and state for each level, for all $256$ rules.

	\subsection*{Coordination number}

	The coordination number is the mean number of springs extending from each node, and is conventionally denoted $\langle z \rangle$. To calculate the coordination number of the generated objects, the coordinates of the smallest triangles (the final level) are treated as nodes, and the edges as springs. To compute this, the number of triangles with state 1 of the last level are counted. Then from these triangles, unique nodes and edges are counted. The coordination number can then be calculated by dividing the number of unique edges by the number of unique nodes as 
	\begin{equation}
		\langle z \rangle=\frac{\textrm {$2 \times $ no of edges}}{\textrm{no of nodes}}\:.
		\label{e:coord}
	\end{equation}
	\newline

	\subsection*{Fractal dimension}

	Fractal objects cannot be described by standard quantities such as area or volume, instead they are described by fractal dimension. The dimension of a fractal such as the Sierpinski triangle is suggested to lie somewhere between 1 and 2~\cite{Fractals}. For the Sierpinski triangle it is known that the fractal dimension is equal to $\frac{\log {3}}{\log {2}} \approx 1.58$~\cite{FractalGeomNature}. Hence, calculating the fractal dimension of all generation rules can lead to the discovery of new fractals beyond the Sierpinski triangle. To compute the fractal dimension, a method referred as similarity dimension, which is applicable only for self-similar sets like the generalized Sierpinski triangle, was used~\cite{FractalGeometry}. This method consists of counting the number of smallest triangles with state 1, $N$, and calculating the factor that the length of the initial triangle has been decreased by. The fractal dimension $d_{\rm f}$ is then given by
	\begin{equation}
		d_{\rm f}=-\frac{\log {N}}{\log {r}}\:,
		\label{f:df}
	\end{equation}  
	where m is the number of triangles with state 1, and r is the edge length of the smallest triangle given by
	\begin{equation}
		r=\frac{1}{2^{m}}
	\end{equation}
	for $m$ levels of recursion, when the initial (largest) triangle has an edge length of~1.

	\subsection*{Viscoelastic spectrum $G^{*}(\omega)$} 
	
	The calculation of the linear viscoelastic spectrum $G^{*}(\omega)$ for spring networks has been detailed previously~\cite{SpringNetworks}. In brief, a series of upright and inverted triangles are generated from the rules as previously defined for $m$ levels of recursion, and then tiled into a rectangular periodic box. The edges of small triangles are mapped to springs with a uniform spring constant $k$. Neglecting inertia, the total force on each node, being the sum of all spring forces plus a drag term deriving from the implicit solvent, must be zero. Note there are no hydrodynamic interactions; this is strictly the Rouse limit~\cite{DoiEdwards}. To allow mapping to a matrix equation for efficient solution, the displacements for node $\alpha$ are first written in the form ${\bf u}^{\alpha}(t)={\bf u}^{\alpha}_{\omega}e^{i\omega t}$, and then the force balance equations are linearised in the complex amplitudes ${\bf u}_{\omega}$,
		\begin{equation}
			\zeta\left(
				{\bf v}^{\alpha}_{\rm w}-i\omega{\bf u}^{\alpha}_{\omega}
			\right)	
			=
			\sum_{\beta\in N(\alpha)}
			k\left\{
				\left[
					{\bf u}^{\beta}_{\omega}-{\bf u}^{\alpha}_{\omega}
				\right]\cdot\hat{\bf t}^{\alpha\beta}
			\right\}
			\hat{\bf t}^{\alpha\beta}\:,
			\label{e:fibreNetHI}
		\end{equation}
		where $\zeta$ is the damping coefficient, the sum is over all nodes $\beta$ directly connected to $\alpha$ by a spring, and $\hat{\bf t}^{\alpha\beta}$ is the unit vector from $\alpha$ to $\beta$. The affine fluid velocity ${\bf v}^{\alpha}_{\rm w}$ is known for each node position. The single node equations (\ref{e:fibreNetHI}) are assembled into a single sparse matrix and solved, and the resulting spring forces converted to $G^{*}(\omega)$ using standard methods; see~\cite{SpringNetworks}. Note that only a single matrix solve is required for each frequency $\omega$, for each network realisation.

	%
	%
	\section*{Results} 
	All $2^{8}=256$ generation rules were systematically varied over a range of  recursive levels. Those for which the fractal dimension was clearly approaching 0, 1 or 2 as the number of levels increased were discarded as being non-fractal. This process left 4 rules that produced fractal geometries, with one being the standard Sierpinski triangle. The fractal dimension and coordination for each are given in Table~\ref{t:fourFracs}.

	\begin{table}[htbp]
		\begin{center}
			
			\begin{tabular}{l|l|l} 
				\textbf{(T0,L0,R0,M0,T1,L1,R1,M1)} & \textbf{$d_{\rm f}$ } & \textbf{ $\langle z \rangle$}\\
				\hline
				(0,0,0,0,1,1,1,0) - Sierpinski triangle & 1.585 & 3.999\\
				(0,0,0,0,0,1,1,1) & 1.585 & 3.999\\
				(0,0,0,0,1,0,1,1) & 1.585 & 3.999\\
				(0,0,0,0,1,1,0,1) & 1.585 & 3.999\\
			\end{tabular}
		\end{center}
		\caption{The fractal dimension $d_{\rm f}$ and mean coordination number $\langle z\rangle$ for the four non--trivial rules uncovered by our systematic exploration of all possible rules.}
		\label{t:fourFracs}
	\end{table}

\begin{figure}[htbp]
		\centering
		\begin{subfigure}[h]{0.25\textwidth}
			\centering
			\includegraphics[scale=0.51,page=1]{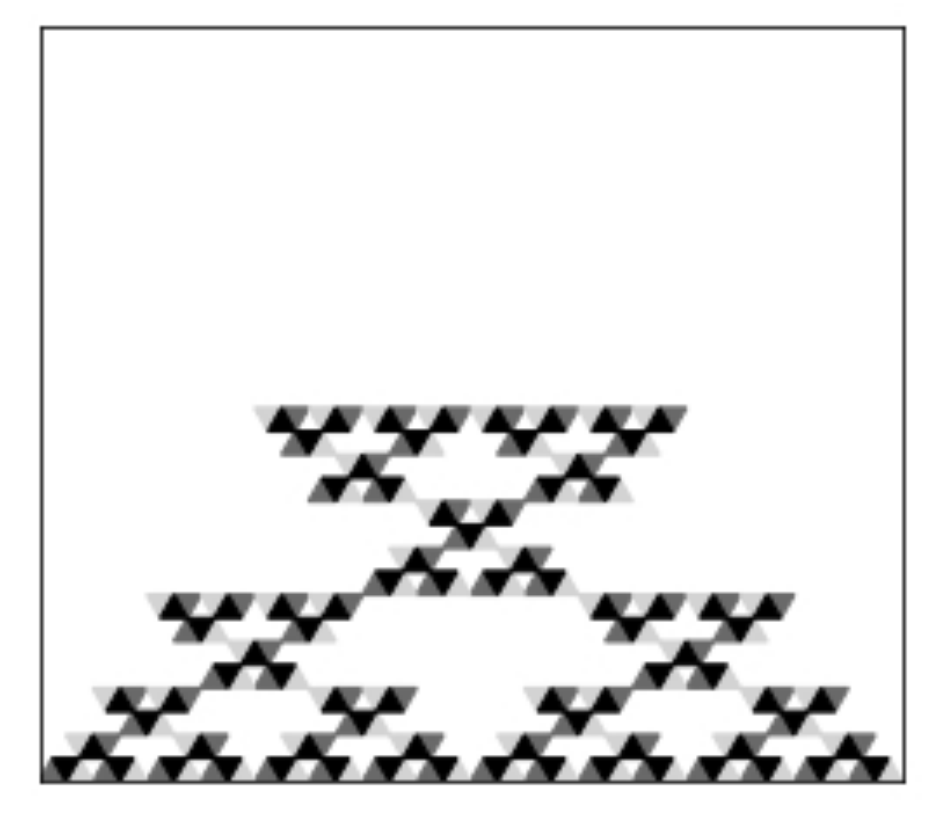}
			\caption{T1=0}
			
		\end{subfigure}%
		~ 
		\begin{subfigure}[h]{0.25\textwidth}
			\centering
			\includegraphics[scale=0.51,page=2]{newfractals.pdf}
			\caption{L1=0}
			
		\end{subfigure}
		~ 
		\begin{subfigure}[h]{0.25\textwidth}
			\centering
			\includegraphics[scale=0.51,page=3]{newfractals.pdf}
			\caption{R1=0}
			
		\end{subfigure}
		\caption{Figures of the three new fractal geometries, plotted for level 5 with sub-triangles having different colour depending on their type (T,L,R,M).}
		\label{fig:3}
		
	\end{figure}

	The three new fractals shown in Fig.~\ref{fig:3} have the same fractal dimension $d_{\rm f}$ and coordination number $\langle z\rangle$ as the Sierpinski triangle. The $\langle z\rangle$ is very close to the marginally rigid value $4$ for spring networks in 2D~\cite{SpringNetworks}, thus the new fractals are also marginally rigid. The fractal dimension for the new fractals can also be calculated analytically. Given that all rules have $T0=L0=R0=M0=0$, the number of triangles with state 1 after $m$ levels of recursion is given by
	\begin{equation}
		N=(T1+L1+R1+M1)^{m}
	\end{equation}
	In the case of the three new fractals, $N=3^{m}$ since all rules have 3 triangles with state 1 and one triangle with state 0. Given equation (\ref{f:df}), the fractal dimension can be calculated as
	\begin{eqnarray}
		 d_{\rm f}&=&-\frac{\log{3^{m}}}{\log{\frac{1}{2^{m}}}}\\
		 &=&\frac{\log{3^{m}}}{\log{2^{m}}}\\
		 &=&\frac{\log{3}}{\log{2}}\approx 1.58\:.
	\end{eqnarray}
	Hence, the generated fractal objects and the Sierpinski triangle are related and share the same statistical properties, but may have different mechanical properties.
	
	We note that the same argument can be used to show that $d_{\rm f}=1$ when exactly two of $T1$, $L1$, $R1$ and $M1$ are unity, still with $T0=L0=R0=M0=0$, since now $N=2^{m}$ and hence $d_{\rm f}=\frac{\log 2}{\log 2}=1$. For rules with $b=T0+L0+R0+M0>0$, a general equation for $N$ can be derived from which the fractal dimension can be inferred. Using $a=T1+L1+R1+M1$, the number of filled triangles after $m$ levels of recursion, $N_{m}$, is related to the number at the previous level by $N_{m}=aN_{m-1}+b(4^{m-1}-N_{m-1})$, where $4^{m-1}-N_{m-1}$ is the number of empty triangles at level $m-1$. This can be shown by induction to obey $N_{m}=a^{m}+b\sum_{i=1}^{m-1}\left(4^{i}-a^{i}\right)\left(a-b\right)^{m-i-1}$. Rearranging the summation into two finite sums and evaluating gives, after a little algebra,
	\begin{equation}
		N_{\rm m}= \frac{1}{4+b-a}\left[4^{m}b + (4-a)(a-b)^m\right]	
		\:.
		\label{e:Nm}
	\end{equation}
	For $b=0$ we recover $N_{m}=a^{m}$ as previously described. For $b>0$, the term $\propto 4^{m}$ will dominate for large $m$. Given equation (\ref{f:df}), this means that the fractal dimension will approach $d_f\rightarrow \log 4/\log 2=2$ for large $m$ for all non-trivial rules. Although an integer $d_{\rm f}$ does not rule out a fractal object exhibiting scale invariance~\cite{FractalGeometry}, visual inspection of the geometries generated by these rules confirm they do indeed generate simple geometries that are non--fractal. These rules were therefore not considered further.
	\onecolumngrid
	
	\twocolumngrid
	
\begin{figure}[htbp]
		\begin{center}
			\includegraphics[width=8.6cm]{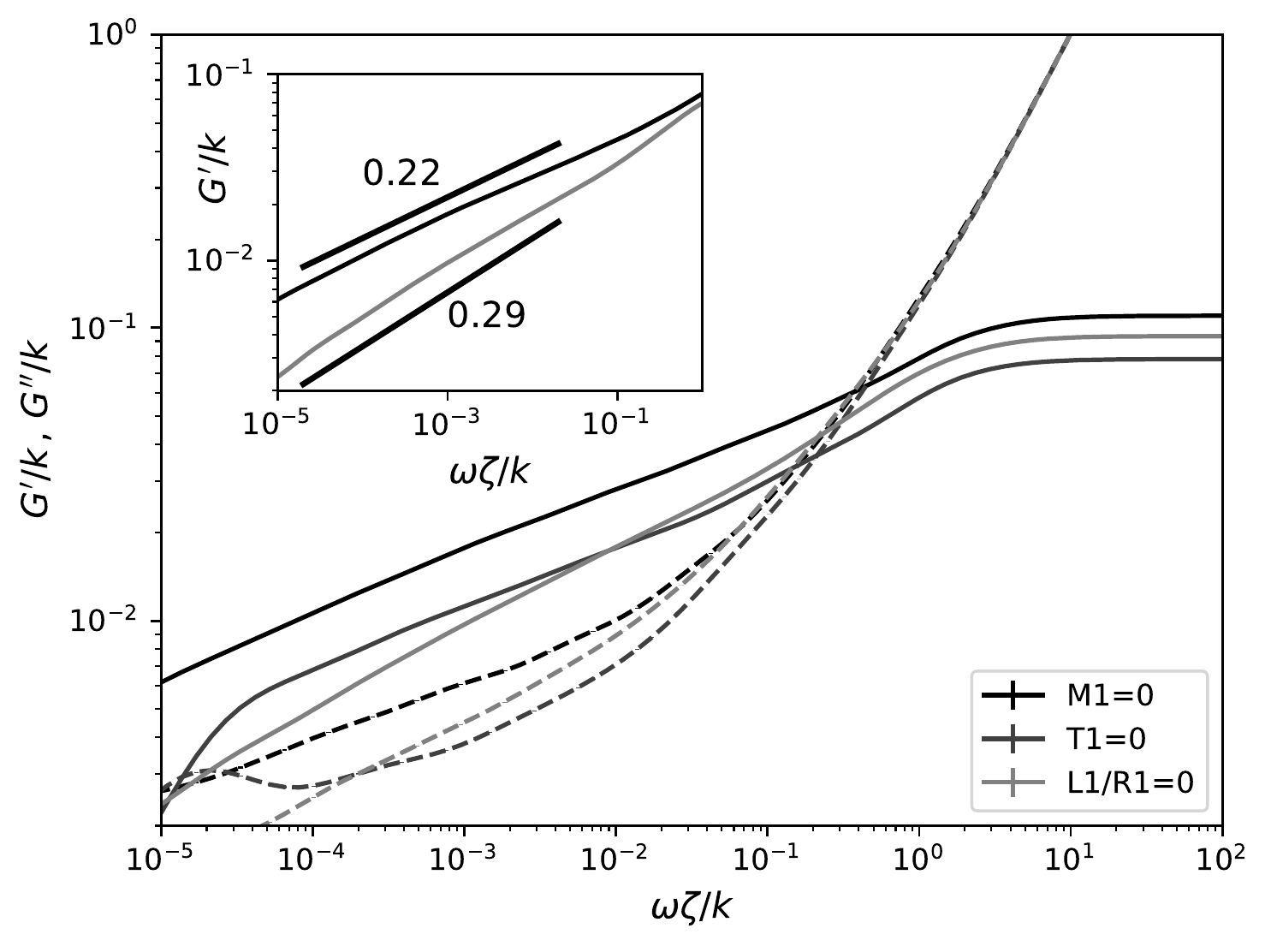}
		\end{center}
		\caption{
			\emph{(Main)} The storage $G^{\prime}(\omega)$ (solid lines) and loss $G^{\prime\prime}(\omega)$ (dashed lines) moduli for the generalized Sierpinski triangles given in the legend, where the label shows which of the four rules $T1$, $L1$, $R1$ or $M1$ are zero. All examples had $m=7$ levels of recursion and data has converged with system size for the frequency range shown.
			\emph{(Inset)} A close--up of $G^{\prime}(\omega)$ for two $M1=0$ and $L1=0$. The straight line segments have the given slope.
		}
		\label{f:visco}	
	\end{figure}

	The viscoelastic spectra $G^{*}(\omega)$ for the spring networks generated by the rules with all of the $T1$, $L1$, $R1$ and $M1$ equal to 1 except exactly one that is equal to zero, including the standard Sierpinski triangle ($M1=0$), are shown in Fig.~\ref{f:visco}. Note that the networks generated by $L1=0$ and $R1=0$ exhibit identical response as they are equivalent with respect to the direction of the applied shear. The networks generated from $T1=0$, $L1=0$ and $R1=0$ did not converge with system size for low frequencies, therefore only frequencies for which the data has converged are shown. The previous findings for $M1=0$~\cite{SpringNetworks} are confirmed; that is, $G^{\prime}(\omega)\propto G^{\prime\prime}(\omega)\propto\omega^{\Delta}$. The networks generated by $T1=0$ appear to follow the same trend, with a similar exponent~$\Delta$, until reaching frequencies around $\omega\zeta/k\approx10^{-4}$ when there is an unusual modulation in the spectrum. This is not a finite size effect, but we are unable to provide an explanation at present, merely pointing out that, as it only arises for very low frequencies, it will be challenging to observe in experiments.
	
	More interesting are the spectra for $L1=0$ and $R1=0$. These follow the same trend, but apparently with a higher exponent. This is confirmed by the inset to the figure, which shows that $G^{\prime}(\omega)$ for $L1=0$ is clearly different to value $\Delta\approx0.22$ for networks generated from the standard Sierpinski triangle $M1=0$. Instead, the exponent is more consistent with a value closer to $\Delta\approx0.29$. This is not a finite size effect, and does not depend on the number of levels of recursion used to generated the networks. We conclude that the viscoelastic exponent $\Delta$ is different for $L1=0$ and $R1=0$, than for $M1=0$ and $T1=0$, despite having the same fractal dimension $d_{\rm f}$, coordination number $\langle z\rangle$, and solvent conditions. 
	
	%
	%
	\section*{Conclusions} 
	Here we have investigated the viscoelastic properties of spring networks deriving from a class of fractals that generalizes the Sierpinski triangle. After numerically producing all $256$ generation rules, the generated objects were analysed based on their geometry, i.e. their fractal dimension and coordination number. From the calculated fractal dimension the rules that produced non-fractal objects were discarded (fractal dimension approaching 0, 1 or 2) and the rules producing fractal objects were then analysed based on their mechanical properties. Four generation rules were found to have fractal geometry, one of which being the standard Sierpinski triangle. These four fractals share the same statistical properties, coordination number $\langle z\rangle\approx{4}$ and fractal dimension $d_{\rm f}\approx{1.58}$. However, when the viscoelastic spectrum of the four fractals was calculated it was shown that despite having the same fractal dimension $d_{\rm f}$ and coordination number $\langle z\rangle$, two values of $\Delta$ were measured, $\Delta\approx0.29$ (for L1=0 and R1=0 ) as well as the previous value $\Delta\approx0.22$ (for the Sierpinski triangle and T1=0). From this we conclude that $\Delta$ does not purely depend on $d_{\rm f}$, even for networks with fixed connectivity.

	%
	%
	\section*{Acknowledgements}
	This work was funded by the University of Leeds, UK.

	\bibliography{bibliography} 
\end{document}